\documentclass[preprint,proceedings]{rmaa}
\usepackage{paralist}

\usepackage{psfrag,color}

\newcommand{\grsim}{\mathrel{\hbox{\rlap{\hbox{\lower4pt\hbox{$\sim$}}}\hbox{$>$}}}}

\SetYear{2003}
\SetConfTitle{Energetics of Cosmic Plasmas}

\title{Physical conditions in NGC 6543 from ISO and HST data \\*[0.3cm] 
and the pitfalls of temperature maps}

\author{
  V. Luridiana\altaffilmark{1}, 
  E. P\'erez\altaffilmark{1},
  and M. Cervi\~no\altaffilmark{1,2}
}

\altaffiltext{1}{Instituto de Astrof\'\i sica de Andaluc\'\i a (CSIC), Granada, Spain.}
\altaffiltext{2}{Laboratorio de Astrof\'\i sica Espacial y F\'\i sica Fundamental, Madrid, Spain.}

\shortauthor{Luridiana et al.}
\shorttitle{Physical conditions in NGC 6543}

\fulladdresses{
\item V. Luridiana: Instituto de Astrof\'\i sica de Andaluc\'\i a (CSIC), 
c/ Camino Bajo de Hu\'etor 24, Apartado 3004, 18080 Granada, Spain
  (\email{vale@iaa.es}).
\item E. P\'erez: Instituto de Astrof\'\i sica de Andaluc\'\i a (CSIC), 
c/ Camino Bajo de Hu\'etor 24, Apartado 3004, 18080 Granada, Spain
  (\email{eperez@iaa.es}).
\item M. Cervi\~no: Instituto de Astrof\'\i sica de Andaluc\'\i a (CSIC), 
c/ Camino Bajo de Hu\'etor 24, Apartado 3004, 18080 Granada, Spain
  (\email{mcs@laeff.esa.es}).
}

\listofauthors{V. Luridiana, E. P\'erez, \& M. Cervi\~no}

\indexauthor{Luridiana, V.}
\indexauthor{P\'erez, E.}
\indexauthor{Cervi\~no, M.}

\abstract{We determine the physical conditions in {NGC~6543},
obtaining two different estimates of the electron temperature ($T_e$)
and one estimate of the electron density ($N_e$).
The electron temperature is computed by means of 
the [O{\sc~iii}] ratio $\lambda\,5007$/$\lambda\,4363$, 
and of the diagnostic diagram combining $\lambda\,5007$ to 
the [O{\sc~iii}] infrared lines $52\,\mu$ and $88\,\mu$. 
The continuum intensity measured on slit spectra is much higher than
theoretically predicted under the simplest assumptions.
After considering several possibilities, 
we suggest enhanced 2-photon emission 
as the most probable source of the additional continuum.
While $T_e$ and $N_e$ derived from the diagnostic diagram
agree with the most recent determination,
$T_e$ derived from $\lambda\,5007$/$\lambda\,4363$ 
is smaller than previously published values, probably due to the bias
in $\lambda\,4363$ introduced by the uncertainty in the continuum.
Our main conclusion is that it is not possible, with present-day data,
to derive accurate temperature maps of photoionized nebulae.}

\resumen{Determinamos las condiciones f\'\i sicas en 
NGC~6543, obteniendo dos distintas estimaciones
de la temperatura electr\'onica ($T_e$) y una de la densidad electr\'onica ($N_e$).
La temperatura electr\'onica es calculada por medio del cociente
de [O{\sc~iii}] $\lambda\,5007$/$\lambda\,4363$, y del diagrama de diagn\'ostico 
que combina $\lambda\,5007$ con las l\'\i neas infrarrojas de [O{\sc~iii}] 
$52\,\mu$ y $88\,\mu$. 
La intensidad en el continuo medida en espectros \'opticos es mucho
mayor que la predicha bajo las hip\'otetis m\'as sencillas.
Sugerimos que la fuente m\'as probable del continuo adicional
es la conversi\'on de fotones Ly$\alpha$ a continuo de dos fotones.
Aunque los valores de $T_e$ y $N_e$ derivados con el diagrama de diagn\'ostico 
est\'an en acuerdo con las determinaciones m\'as recientes, 
$T_e$ derivada del cociente $\lambda\,5007$/$\lambda\,4363$
es inferior a los valores publicados en la literatura,
probablemente debido al sesgo en $\lambda\,4363$ introducido por la
incertidumbre en el continuo.
Nuestra conclusi\'on principal es que no es posible, 
con los datos actualmente disponibles, 
obtener mapas precisos de temperatura de nebulosas fotoionizadas.
}

\addkeyword{instrumentation: miscellaneous}
\addkeyword{planetary nebulae: general}
\addkeyword{planetary nebulae: individual: NGC 6543}

\begin{document}
\maketitle

\section{Introduction}\label{sec:intro}

In this work, we investigate the physical conditions
in the planetary nebula NGC 6543
using two different diagnostics:
the temperature-sensitive 
[O{\sc~iii}] ratio $I(\lambda\,5007)/I(\lambda\,4363)$,
and the ($T_e$, $N_e$) diagnostic diagram
based on the [O{\sc~iii}] optical and infrared lines
proposed by \citet{DLW85} (henceforth DLW).
The intensity of the infrared lines is obtained from ISO archive spectra,
and the intensity of [O{\sc~iii}] $\lambda\lambda\,5007,\,4363$ 
from HST archive images.
Additionally, H$\alpha$ and H$\beta$ HST images 
were used to determine the reddening correction to be applied to the other images,
and HST images around $\lambda\, 6584$ were used to correct the H$\alpha$ images for the 
[N{\sc~ii}] contribution.
Finally, long-slit data obtained with the 2.5m 
Isaac Newton Telescope (INT) at the Observatorio del Roque de los
Muchachos, on La Palma,
were used to correct the HST images 
for the contribution of other lines and the continuum.
A full description of this research can be found in
\citet*{LPC03}.

\section{The ISO data}\label{sec:iso}

NGC 6543 (the `Cat's eye') 
was routinely observed by ISO for calibration purposes,
thereby a large set of spectra is available.
We selected 20 out of the 92 Long-Wavelength Spectrometer 
spectra available in the ISO archive, 
and based our analysis on the data
measured with the SW2 and LW1 detectors only, 
obtaining the following line intensities:
$I(52\mu)=(5.10\pm 0.61)\times 10^{-10}\ {\rm erg\,sec}^{-1} {\rm cm}^{-2}$,
and
$I(88\mu)=(1.39\pm 0.17)\times 10^{-10}\ {\rm erg\,sec}^{-1} {\rm cm}^{-2}$.

\section{The INT data}\label{sec:int}

\begin{figure}[!t]
 \includegraphics[width=\columnwidth]{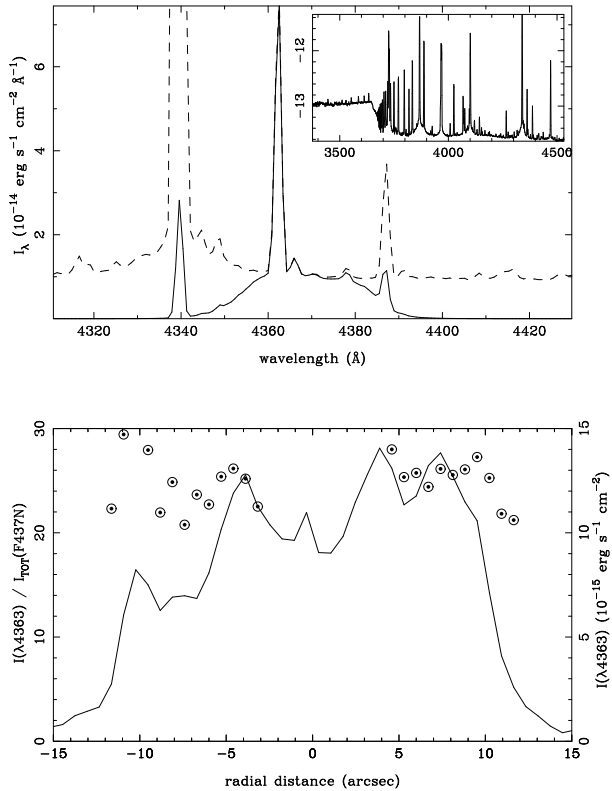}
 \caption{Top: nebular spectrum of NGC 6543 around $\lambda\,4363$,
along position angle 5$^{\circ}$.
Bottom, left handside scale: the dotted circles show the percent of 
the total intensity in F437N contributed by the $\lambda\,4363$ line;
right hand side scale: variation of $I(\lambda\,4363)$ along 
the slit.\label{fig:spec}}
\end{figure}

NGC 6543 was observed spectroscopically as part of a wider study of PNe
on 1995 July 8 and 9, 
using the Intermediate Dispersion Spectrograph attached to the
INT \citep[see][for further details]{LPC03}.
The top panel of Figure~\ref{fig:spec}
illustrates the very high quality of these data.

\section{The HST images\label{sec:HST}}

From the HST data archive, we retrieved images of NGC~6543
through the narrow band filters F437N, F487N, F502N, 
F656N, and F658N,
centered on the emission lines [O{\sc~iii}] $\lambda\,4363$, H$\beta$,
[O{\sc~iii}] $\lambda\,5007$, 
H$\alpha$, and [N{\sc~ii}] $\lambda\,6584$ respectively. 
The calibration of the images is conceptually straightforward, but the procedure
contains several potential pitfalls. 
One of the most delicate point is the estimation of the contribution 
to the flux of the continuum and of neighbouring lines: see, e.g., \citet{ODD99}.
In the following, we will describe some problems related to
the continuum subtraction.

\subsection{Continuum subtraction}

The processes contributing to the nebular continuum flux are {\sc H~i}, {He\sc~i},
and {He\sc~ii} recombination, bremsstrahlung, and 2-photon decay.
The 2-photon continuum intensity is quite difficult
to compute accurately, as it depends not only on the local physical
conditions $N_e$ and $T_e$, but also on the 
fate of the Ly$\alpha$ photons produced by recombination
\citep{BM70}.
The continuum contribution is very important in filters centered on weak lines,
such as [{\sc O~iii}] $\lambda\,4363$,
but it barely affects the intensity in filters centered on
strong lines, such as H$\alpha$ or [{\sc O~iii}] $\lambda\,5007$.
Our computations show that, neglecting Ly$\alpha$ conversion,
the continuum yields 40 percent of the total intensity 
in F437N. Subtracting out the contribution of the other lines,
we found that $\lambda\,4363$ yields 44 percent of the total flux.

We compared this computation with a measurement of
the continuum flux on the INT spectra,
finding that the continuum contributes 58 percent
and that $\lambda\,4363$ contributes only 26 percent
of the total intensity in F437N (Figure~\ref{fig:spec}, bottom).
To investigate the discrepancy with respect 
to our previous calculations, we considered
three likely mechanisms that would enhance the continuum:
X-ray emission, dust scattering, and 2-photon emission.

\begin{figure}[!b]
 \includegraphics[angle=-90,width=\columnwidth]{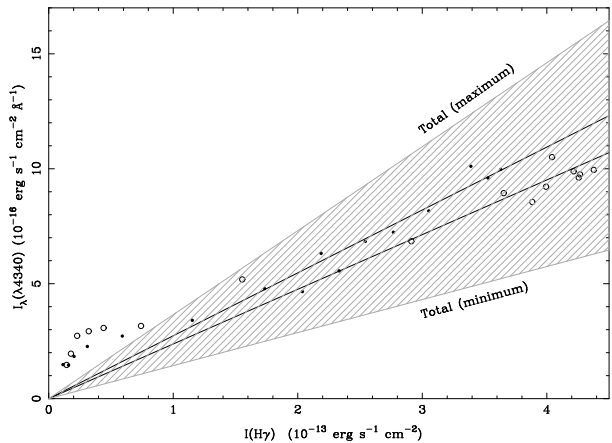}
 \caption{Observed H$\gamma$ and nearby continuum fluxes
measured along the INT slit, at different distances 
from the central star. The shaded region corresponds to the region
predicted by recombination theory.
Dots and open circles represent the two directions along the slit,
and the two solid lines the corresponding best-fit linear relations.
 \label{fig:contHgamma}}
\end{figure}

The third possibility is the most plausible to explain
the difference between the spectroscopic data and our computations.
The occurrence of Ly$\alpha$ conversion 
can enhance the total continuum in the region around
$\lambda\,4363$ (where 2-photon emission is a dominant process)
by as much as a factor 2.5. 
The detailed calculation of the actual efficiency of this process is virtually
impossible, since it would require knowledge of the local escape probabilities
and the dust structure of the nebula, and the implementation of this information
in a 3D computation. 
The only feasible alternative is to bracket this process
by computing the 2-photon emission in
the extreme cases of minimum and maximum efficiencies.
Figure~\ref{fig:contHgamma} compares the observational data to the continuum range
predicted by recombination theory,
as a function of the H$\gamma$ intensity. The data are
compatible with a Ly$\alpha$-conversion enhancement of the 2-photon continuum.
The figure also shows that the relation 
between H$\gamma$ and the measured continuum is not linear,
and that there are large fluctuations around the best-fit straight line:
this agrees with our hypothesis,
since the Ly$\alpha$ conversion process acts with different
efficiencies across the nebula.
This fact implies
that it is not possible to rely on theoretical calculations alone
to correct the total flux in each image for the continuum contribution,
and that precision photometry of weak lines cannot
be done with HST/WFPC2 images in the absence of data
specifically designed to measure the continuum.

\begin{figure}[!t]
 \includegraphics[width=\columnwidth]{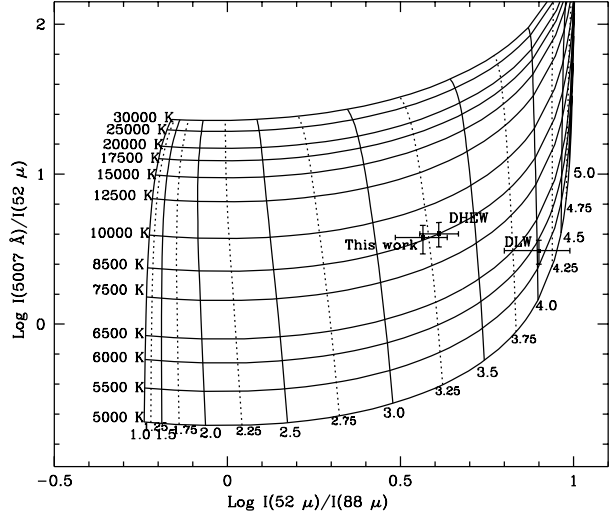}
 \caption{[O{\sc~iii}] infrared-line diagnostic diagram.
 \label{fig:OIII}}
\end{figure}

To carry on our analysis in spite of these difficulties, 
we adopted the continuum flux measured on the INT spectra.
Combining the infrared intensities measured on ISO spectra
with the [{\sc O~iii}] $\lambda\,5007$ intensity
derived from the HST data, we obtained the
point shown in the diagnostic diagram of Figure \ref{fig:OIII}, 
where the data points by DLW and \citet{Dal95} (DHEW) are also plotted.
This figure shows that our results disagree with 
the point by DLW, and are in excellent agreement with those by DHEW.
The values we obtain are $N_e = 1650^{+550}_{-400}$ cm$^{-3}$, $T_e = 8600 \pm 500$ K,
while the corresponding values quoted by DLW and DHEW are 
$N_e$= $10000^{+\infty}_{-6000}$ cm$^{-3}$, $T_e = 5800 \pm 300$ K
and $N_e$= $2000^{+500}_{-400}$ cm$^{-3}$, $T_e = 8500 \pm 500$ K
respectively.

\section{Summary and conclusions}

In this work we attempted to derive 
self-consistently  the physical conditions in the bright core of NGC 6543,
using the standard nebular-to-auroral temperature diagnostic,
and the diagnostic diagram based on
infrared lines developed by DLW.
The $N_e$ and $T_e$ values derived by means of the diagnostic diagram
are not compatible with those by DLW,
but they are in very good agreement to the more recent result by DHEW.
On the other hand,
the nebular-to-auroral temperature we derive is somewhat lower than 
the values published in the literature. 
This disagreement may depend on a bias in the 
adopted continuum level, since we found that the continuum level
measured on slit spectra differs from the one expected 
at $T_e \sim T_e($O$^{++})_{opt}$ when Ly$\alpha$ conversion is
neglected.
We investigated several possibilities to explain the extra continuum:
the most plausible turned out to be enhanced 2-photon emission
originated by conversion of scattered Ly$\alpha$ photons,
but we were not able to work out a theoretical prescription
to compute accurately the continuum intensity that would 
eliminate the necessity of relying on spectroscopic information.

As the work progressed, it became evident to us
that the archival data we used were not optimized for this particular task.
As a result, the most important source of uncertainty in the determination of the 
optical temperature is the continuum subtraction in the $\lambda\,4363$ image;
a specific conclusion we draw 
is that it is not possible with these data to obtain an accurate bidimensional
temperature map of the nebula.

\acknowledgments

VL is supported by a Marie Curie Fellowship
of the European Community programme {\sl ``Improving Human Research Potential 
and the Socio-economic Knowledge Base''} under contract number HPMF-CT-2000-00949.
This project has been partially supported by the AYA 3939-C03-01 program.

\end{document}